\input{aipcheck}

\documentclass[
    ,final            
  ]
  {aipproc}

\layoutstyle{6x9}
\usepackage{xspace}

\def\gcm2{g/cm$^2$}

\def\knee{{\sl knee}\xspace}

\def\fref#1{Fig.~\ref{#1}\xspace}

\def\uka{Institut f\"ur Experimentelle Kernphysik,
         Universit\"at Karlsruhe, 76021 Karlsruhe, Germany}
\def\fzk{Institut\ f\"ur Kernphysik, Forschungszentrum Karlsruhe,
         76021~Karlsruhe, Germany}
\def\buc{National Institute of Physics and Nuclear Engineering,
         7690~Bucharest, Romania}

\def\sol{Soltan Institute for Nuclear Studies, 90950~Lodz, 
         Poland}
\def\yer{Cosmic Ray Division, Yerevan Physics Institute, Yerevan 36, Armenia}

\begin{document}

\title{Galactic cosmic rays and the knee - Results from the KASCADE experiment%
      \footnote{Invited talk, Astrophysical Sources of High Energy
      Particles and Radiation, Torun, June 20 - 24, 2005 }}


\author{J.R. H\"orandel}{address=\uka}
\author{W.D.~Apel}{address=\fzk}
\author{F.~Badea}{address=\fzk}
\author{K.~Bekk}{address=\fzk}
\author{A.~Bercuci}{address=\buc}
\author{J.~Bl\"umer}{address=\fzk,altaddress=\uka}
\author{H.~Bozdog}{address=\fzk}
\author{I.M.~Brancus}{address=\buc}
\author{A.~Chilingarian}{address=\yer}
\author{K.~Daumiller}{address=\fzk}
\author{P.~Doll}{address=\fzk}
\author{R.~Engel}{address=\fzk}
\author{J.~Engler}{address=\fzk}
\author{H.J.~Gils}{address=\fzk}
\author{R.~Glasstetter}{address=\uka,altaddress={now at University of Wuppertal,
                                                42097 Wuppertal, Germany}}
\author{A.~Haungs}{address=\fzk}
\author{D.~Heck}{address=\fzk}
\author{K.-H.~Kampert}{address=\uka,altaddress=\fzk,altaddress={now at
                       University of Wuppertal, 42097 Wuppertal, Germany}}
\author{H.O.~Klages}{address=\fzk}
\author{G.~Maier}{address=\fzk,altaddress={now at University of Leeds,
                                           LS2~9JT~Leeds, United Kingdom}}
\author{H.J.~Mathes}{address=\fzk}
\author{H.J.~Mayer}{address=\fzk}
\author{J.~Milke}{address=\fzk}
\author{M.~M\"uller}{address=\fzk}
\author{R.~Obenland}{address=\fzk}
\author{J.~Oehlschl\"ager}{address=\fzk}
\author{S.~Ostapchenko}{address=\fzk, altaddress={ on leave of absence from
        Moscow State University, 119899~Moscow, Russia}}
\author{M.~Petcu}{address=\buc}
\author{S.~Plewnia}{address=\fzk}
\author{H.~Rebel}{address=\fzk}
\author{A.~Risse}{address=\sol}
\author{M.~Risse}{address=\fzk}
\author{M.~Roth}{address=\uka}
\author{G.~Schatz}{address=\fzk}
\author{H.~Schieler}{address=\fzk}
\author{H.~Ulrich}{address=\fzk}
\author{J.~van~Buren}{address=\fzk}
\author{A.~Vardanyan}{address=\yer}
\author{A.~Weindl}{address=\fzk}
\author{J.~Wochele}{address=\fzk}
\author{J.~Zabierowski}{address=\sol}

\begin{abstract}
 Results of the KASCADE air shower experiment investigating the origin of
 cosmic rays in the energy region from $10^{13}$ to $10^{17}$~eV are
 presented. Attention is drawn on the investigation of interactions in the
 atmosphere and the energy spectrum and mass composition of cosmic rays.
\end{abstract}

\maketitle


\section{Introduction}

One of the most remarkable structures in the energy spectrum of cosmic rays is
a change of the spectral index $\gamma$ of the power law $dN/dE\propto
E^\gamma$ at an energy of about 4~PeV, the so called \knee
\cite{naganowatson,haungsrebelroth,pg}.  
The origin of the \knee has not been resolved yet, however an explanation of
the \knee is thought to be a corner stone in the understanding of the origin of
galactic cosmic rays.
In the literature various reasons for the origin of the \knee are discussed
related to the acceleration and propagation processes of cosmic rays as well as
to interactions in interstellar space or the Earth's atmosphere
\cite{origin,ecrsreview}.

The steeply falling energy spectrum requires large detection areas at high
energies.  The largest balloon borne experiment with single element resolution
(TRACER, 5~m$^2$~sr) reaches energies of a few $10^{14}$~eV \cite{tracer05}.
To study higher energies experiments covering several $10^4$~m$^2$ and exposure
times exceeding several years are necessary, which, at present, can only be
realized in ground based installations. They measure the secondary products
generated by high-energy cosmic-ray particles in the atmosphere -- the
extensive air showers.  The challenge of these investigations is to reveal the
properties of the shower inducing particle behind an absorber -- the atmosphere
-- with a total thickness at sea level corresponding to 11 hadronic interaction
lengths or 30 radiation lengths.

One of the most advanced experiments in the energy range from 
$10^{13}$~eV to $10^{17}$~eV is the experiment KASCADE ("KArlsruhe
Shower Core and Array DEtector") \cite{kascadenim}.  It is continuously
operating since 1996, detecting simultaneously the three main components of air
showers.  A $200\times 200$~m$^2$ scintillator array measures the
electromagnetic and muonic components ($E_\mu>0.23$~GeV).  The central detector
system combines a large hadron calorimeter, measuring the energy, as well as
point and angle of incidence for hadrons with energies $E_h>50$~GeV
\cite{kalonim}, with several muon detection systems ($E_\mu>0.49$, 2.4~GeV)
\cite{mwpcnim}.  In addition, high-energy muons are measured by an underground
muon tracking detector equipped with limited streamer tubes ($E_\mu>0.8$~GeV)
\cite{mtdnim}.
The multi-detector set-up allows to address questions connected to the
development of air showers in the atmosphere and the origin of high energy
cosmic rays. 

\section{High-energy interactions}

Addressing astrophysical questions with air-shower data necessitates the
understanding of high-energy interactions in the atmosphere.  Or, in reversion,
the interpretation of properties of primary radiation derived from air-shower
measurements depend on the understanding of the complex processes during the
development of air showers.  Recent investigations indicate inconsistencies in
the interpretation of air shower data \cite{rothnn,chicagoknee,pg}.  Thus, one
of the goals of KASCADE is to investigate high-energy interactions and to
improve contemporary models to describe such processes.

For air shower interpretation the understanding of multi-particle production in
hadronic interactions with a small momentum transfer is essential
\cite{engelisvhecripylos}.  Due to the energy dependence of the coupling
constant $\alpha_s$ soft interactions cannot be calculated within QCD using
perturbation theory. Instead, phenomenological approaches have been introduced
in different models. These models are the main source of uncertainties in
simulation codes to calculate the development of extensive air showers, such as
the program CORSIKA \cite{corsika}.  Several codes to describe hadronic
interactions at low energies (GHEISHA, FLUKA, and UrQMD) as well as high
energies (DPMJET, NEXUS, QGSJET, SIBYLL, and VENUS) have been embedded in
CORSIKA. Their predictions are compared with experimental results in order to
check their correctness.

Studies of the shower development in the atmosphere have been performed with
the multi-detector set-up, see e.g. \cite{isvhecri02wwtest}.  For example,
deficiencies of some models have been pointed out \cite{wwtestjpg}, the lateral
distributions of the electromagnetic, muonic, and hadronic components have been
measured \cite{kascadelateral}, and the attenuation and absorption lengths of
showers have been determined \cite{kascadeabslength}.  

Of particular importance for the interaction studies is the large hadron
calorimeter, which has been calibrated recently at an accelerator beam with
hadrons up to 350~GeV \cite{kalocernws}.  It enables studies of the hadronic
backbone of a shower.  Interactions with small momentum transfer have been
studied \cite{rissejpg}. Such interactions are hardly accessible at collider
experiments (only at low energies in fixed target set-ups) and air-shower
experiments provide complementary information.  The arrival time of hadrons in
air showers has been studied \cite{hadrisvhecricern} and attention has been
drawn on the investigation of the geometrical structure of the hadronic shower
core to search for interactions with unusual large transverse momenta
\cite{annaprd}.  The transverse momentum of secondary pions in hadronic
interactions is investigated by the reconstruction of the pseudo rapidity
distributions of high-energy muons registered with the muon tracking detector
\cite{zabipylos}.

A valuable tool to test high-energy interaction models are correlations between
different shower components.  Since the mass composition of cosmic rays is not
well known in the PeV region, in such investigations frequently the
measurements are compared to predictions for proton and iron induced showers as
extreme assumptions for the composition.  Such correlations are, for example,
the dependence of the number of hadrons, the hadronic energy sum, or the energy
of the most energetic hadron on the number of muons or electrons
\cite{jenskrakow,jenspune}.  A couple of years ago some models, like
SIBYLL~1.6, DPMJET~2.5, or NEXUS~2 failed to describe the measurements in
particular correlations.  While  on the other hand, for contemporary models,
like QGSJET~01, SIBYLL~2.1, or DPMJET~2.55 the KASCADE measurements of various
correlations of the electromagnetic, muonic, and hadronic components are
bracketed by the extreme assumptions of primary protons and iron nuclei
\cite{jenskrakow,jenspune}.

\begin{figure}
 \includegraphics[height=65mm]{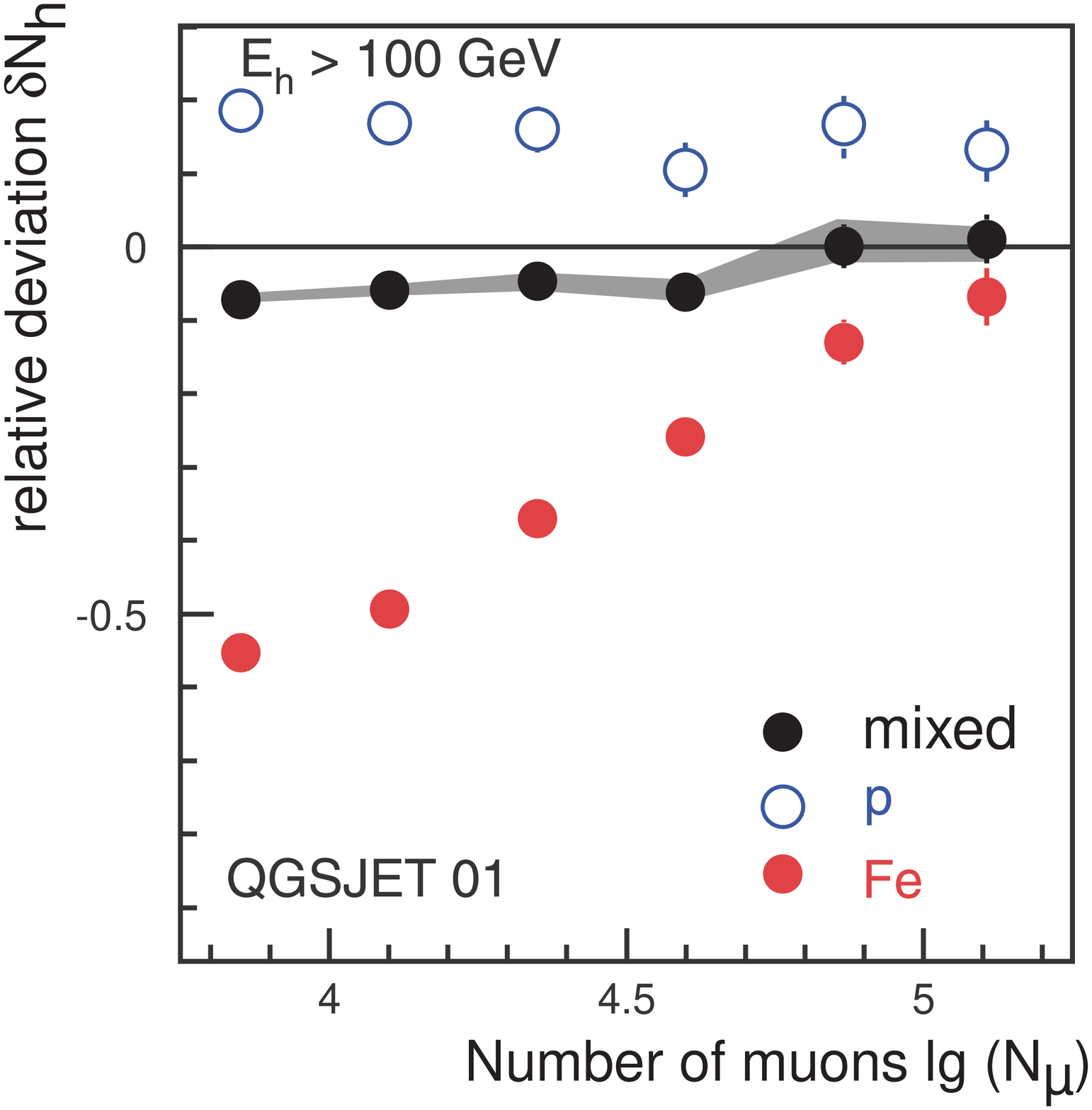}\hspace*{\fill}
 \includegraphics[height=67mm]{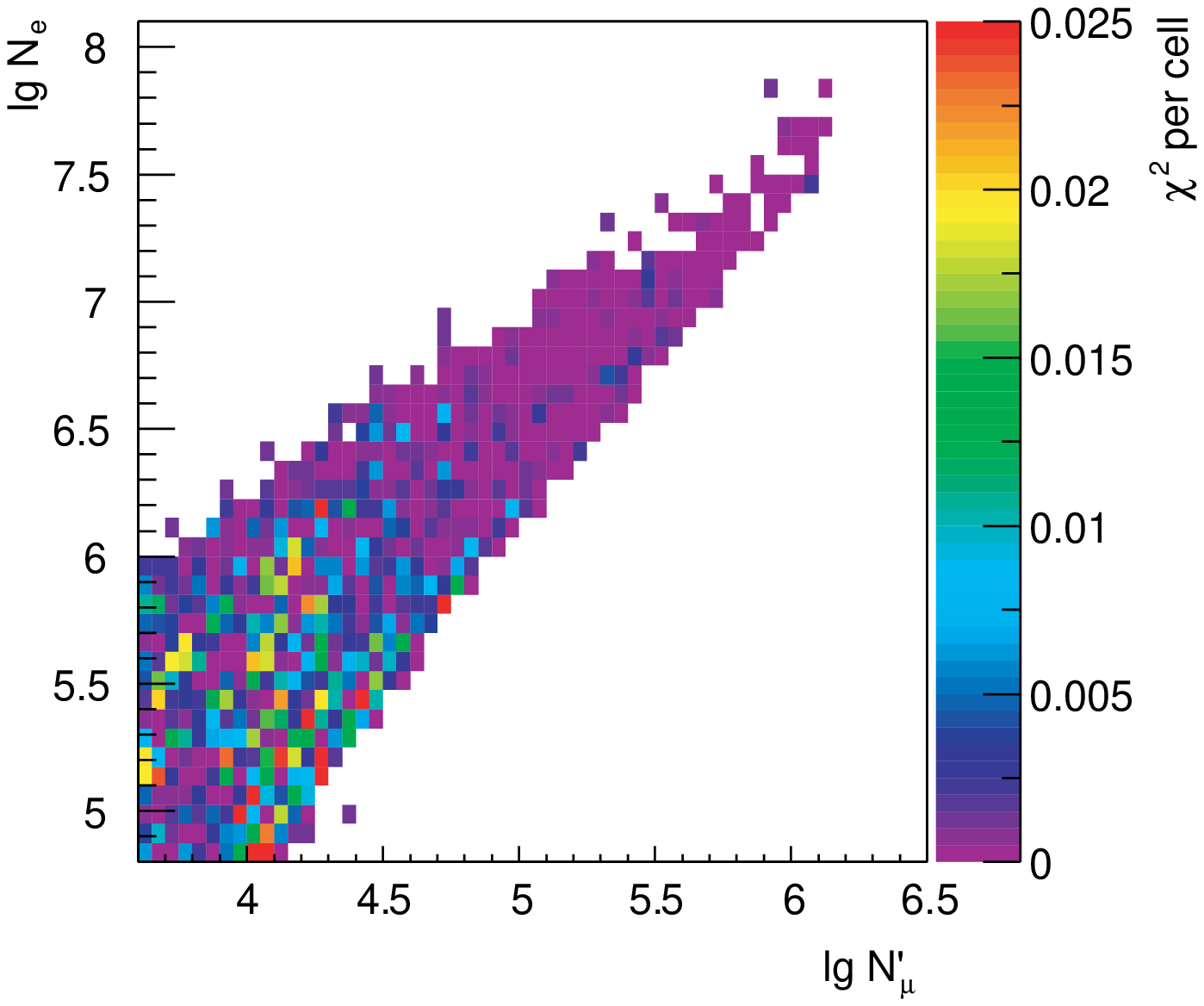}
 \caption{{\sl Left:} Relative deviation of the number of reconstructed
	   hadrons relative to the KASCADE measurements as function of the
	   number of muons for simulations using CORSIKA/QGSJET~01, assuming
	   pure protons and iron nuclei as well as a mixed composition
	   \cite{jenspune}.
	  {\sl Right:} Deviation of the reconstructed and observed
	   two-dimensional shower size spectrum using CORSIKA/QGSJET~01 to
	   interpret the measurements, see text and \cite{ulrichapp}.}
  \label{jens}
\end{figure}

While for previous analyses  pure proton or iron compositions have been
assumed, at present, more detailed analyses are performed \cite{jenspune}. They
take into account the spectra for elemental groups as obtained from
investigations of the electromagnetic and muonic components (see below). The
relative deviation of the reconstructed number of hadrons $\delta N_h=
(N_h^{sim}-N_h^{meas})/N_h^{meas}$ for simulated values $N_h^{sim}$ relative to
the KASCADE measurements $N_h^{meas}$ is shown in \fref{jens} ({\sl left}) as
function of the number of muons.  Presented are values using CORSIKA/QGSJET~01
assuming protons, iron nuclei, and a mixed composition according to the five
elemental groups.  For a fully consistent model the values for the mixed
composition are expected to agree with the measurements, i.e. should be at the
zero line. One recognizes deviations of about 10\% for low muon numbers ($\lg
N_\mu<4.8$), while the model yields consistent results for higher energies. It
is remarkable that the same behavior has been observed for an analysis of the
electromagnetic and muonic component \cite{ulrichapp}, see below.

In conclusion, the models QGSJET~01 \cite{qgsjet}, SIBYLL~2.1 \cite{sibyll21},
and DPMJET~2.55 \cite{dpmjet} seem to be the most reliable models to
describe high-energy hadronic interactions.  However, also they are not able to
describe the data completely consistent \cite{ulrichapp}.  This illustrates the
resolution of experiments like KASCADE, which are able to study details of
high-energy interactions in the atmosphere and indicates the progress made in
this field during the last decade.  At the same time, this stimulates new
efforts, with the objective to improve the present interaction models. Examples
are the study of air-shower relevant parameters at accelerators
\cite{meurericrc}, new theoretical concepts included in the QGSJET model
\cite{qgsjet2}, or the investigations of dedicated changes of physical
parameters like the proton-proton cross-section or the inelasticity of hadronic
interactions on air-shower observables \cite{wq,isvhecri04kascadewq}.  The
latter indicate that variations of interaction parameters within the error
bounds of accelerator measurements yield significant and measurable changes in
the air shower development \cite{kascadewqpune}.

\section{Cosmic-ray anisotropy}
Supernova remnants, such as Cassiopeia A, have been observed in electromagnetic
radiation in a wide energy range up to TeV-energies.  Calculations indicate
that the observed multi-wavelength spectra are consistent with the acceleration
of cosmic-ray electrons and hadrons in supernova remnants \cite{berezhko-casa}.
Recent observations by the H.E.S.S. experiment reveal a shell structure of the
supernova remnant RXJ-1713 and an energy spectrum of $\gamma$-rays $\propto
E^{-2.2}$ in agreement with the idea of particle acceleration in a shock front
\cite{hesssnr}. The spectrum extends up to energies of 10~TeV and provides
evidence for the existence of particles with energies beyond 100~TeV at the
shock front that emerged from the supernova explosion.  
The $\gamma$-ray spectrum is expected to extend to much higher energies due to
the production of neutral pions by charged cosmic rays interacting with the
interstellar matter. 

%
Also, of great interest is to study the arrival direction of charged cosmic
rays to search for potential point sources.  The arrival directions of showers
with energies above 0.3~PeV covering a region from $10^\circ$ to $80^\circ$
declination have been investigated with KASCADE \cite{kascade-points}.  No
significant excess has been observed neither for all showers, nor for muon poor
events.  The analysis has been deepened by investigating a narrow band
($\pm1.5^\circ$) around the Galactic plane. Also circular regions around 52
supernova remnants and 10 TeV-$\gamma$-ray sources have been studied.  None of
the searches provided a hint for a point source, neither by taking into account
all events, nor selecting muon-poor showers only.  Upper limits for the fluxes
from point like sources are determined to be around
$10^{-10}$~m$^{-2}$s$^{-1}$. In addition, no clustering of the arrival
direction for showers with primary energies above 80~PeV is visible.

While the search for point sources is related to the investigation of
cosmic-ray acceleration sites, the large scale anisotropy is expected to reveal
properties of the cosmic-ray propagation.  The Rayleigh formalism is applied to
the right ascension distribution of extensive air showers measured by KASCADE
\cite{kascade-aniso}.  No hints of anisotropy are visible in the energy range
from 0.7 to 6~PeV. This accounts for all showers, as well as for subsets
containing showers induced by predominantly light or heavy primary particles.
Upper limits for Rayleigh amplitudes are determined to be between $10 ^{-3}$ at
a primary energy of 0.7 PeV and $10^{-2}$ at 6 PeV. 
The increase of the amplitudes as function of energy is predicted by
calculations using a diffusion model to describe the cosmic-ray propagation in
the Galaxy \cite{candiaaniso}.  This indicates that leakage from the Galaxy
plays an important part during cosmic-ray propagation and most likely, the
leakage is also (partly) responsible for the origin of the \knee. On the other
hand, simple Leaky-Box models describing cosmic-ray propagation seem to be
ruled out by the measurements \cite{kascade-aniso,maierflorenz,aspenphen}.

In addition, the data registered by the KASCADE experiment were searched for
events which might be attributed to primary $\gamma$-rays \cite{schatzgamma}.
Possible $\gamma$-induced events are identified by their low muon to electron
ratios and by the steepness and smoothness of their electron lateral
distributions.  No local enhancement indicative of a point source was observed.
In comparison with previous measurements, the upper limit could be improved and
the range could be extended to higher energies.

\section{Energy spectra and mass composition}

The main objective of KASCADE is to determine the energy spectra and mass
composition of cosmic rays. The problem has been approached from various points
of view.  It could be shown that a \knee exists in all three main shower
components, i.e.  electrons, muons, and hadrons with a \knee position at about
4 to 5~PeV \cite{allknee}.  The primary energy spectrum could be established
based on the electromagnetic and muonic \cite{glasstetterslc} as well as the
hadronic and muonic shower components \cite{hknie}.  Analyses of the
electromagnetic and muonic shower components \cite{weber}, the hadronic and
muonic components \cite{kascadehm}, as well as various combinations of them
\cite{rothnn} indicate an increase of the mean logarithmic mass of cosmic rays
as function of energy in the \knee region.

The longitudinal development of the muonic shower component is studied with the
muon tracking detector of the KASCADE-Grande experiment \cite{buettner}.  The
arrival direction and point of incidence of individual muons are registered and
allow in conjunction with the geometric parameters of the electromagnetic
shower the reconstruction of the average production height of muons. Through
comparison of measured values with predicted production heights from Monte
Carlo calculations, the mean logarithmic mass of cosmic rays is inferred.  In
the energy range from $\sim0.4$ to $\sim30$~PeV a steady increase of the mean
logarithmic mass has been obtained.

Attention has also been drawn on a special class of events, the unaccompanied
hadrons. These originate from small air showers, which have interacted at large
heights and almost the complete shower has been absorbed in the atmosphere, and
only one single hadron has been reconstructed in the calorimeter
\cite{kascadesh}.  Simulations reveal that these events are closely related to
the flux of primary protons. Hence, the measured flux of unaccompanied hadrons
at ground level has been used to derive the spectrum of primary protons based
on simulations with CORSIKA/QGSJET. The resulting flux follows a single power
law in the energy range from 100~GeV to 1~PeV and is compatible with direct
measurements. 

\begin{figure}
 \includegraphics[width=0.49\textwidth]{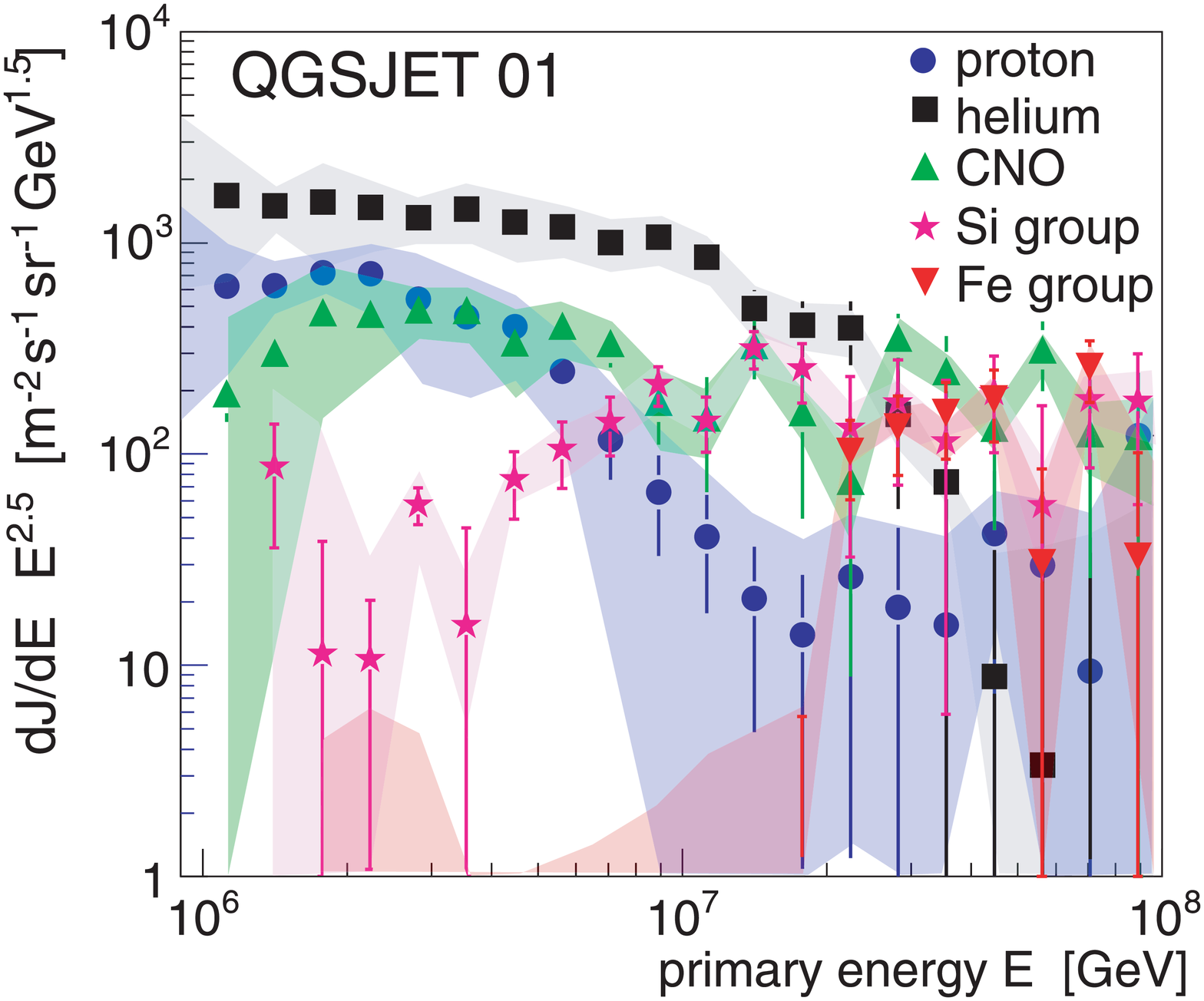}\hspace*{\fill}
 \includegraphics[width=0.49\textwidth]{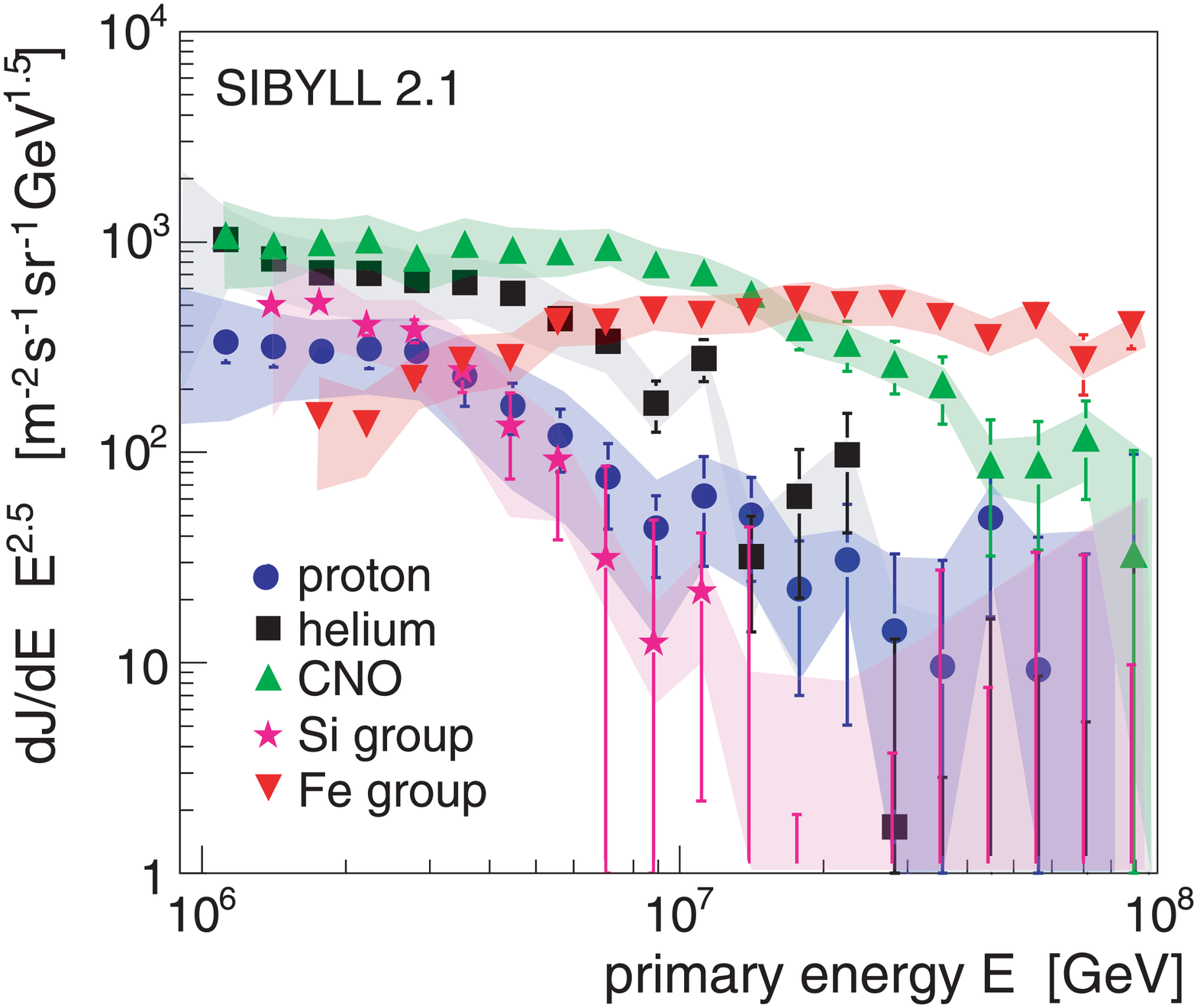}
 \caption{Energy spectra for five elemental groups using the hadronic
	  interaction models QGSJET~01 ({\sl left}) and SIBYLL~2.1 ({\sl
	  right}) to interpret the data \cite{ulrichapp}.}
 \label{holger}	  
\end{figure}

An advanced analysis is founded on the measurement of the electromagnetic and
muonic shower components with the scintillator array \cite{ulrichapp}.  The
analysis is based on the deconvolution of the two-dimensional electron muon
number distribution. The content of each cell in the $N_e$-$N_\mu$ plane is
assumed to be the sum of contributions from five different primary elements,
i.e. protons, helium, carbon, silicon, and iron.  These elements are taken to
represent the full spectrum of cosmic-ray nuclei.  

The application of an unfolding procedure to the data is performed on basis of
two hadronic interaction models (QGSJET~01 and SIBYLL~2.1) as
implemented in the CORSIKA program. The resulting spectra for elemental groups
are displayed in \fref{holger}. The results exhibit sequential cut-offs for the
individual mass groups for the proton, helium, and CNO components.  The
systematic differences for the spectra derived with QGSJET and SIBYLL amount to
a factor of about two to three.  The silicon and iron groups show a rather
unexpected behavior for both models.  The discrepancies are attributed to the
fact that none of the models is able to describe the observed data set in the
whole energy range consistently, for details see \cite{ulrichapp}. 
Assuming the obtained spectra in the simulations, the two dimensional shower
size spectrum has been calculated and compared to the measurements. The
differences in each cell of the $N_e$-N$_\mu$ plane are depicted in \fref{jens}
({\sl right}). One recognizes relatively large deviations at low energies ($\lg
N_\mu<4.8$), indicating inconsistencies for the model QGSJET~01 in this range.
On the other hand, for SIBYLL~2.1 large differences are obtained at high
energies.

Despite of the discrepancies, the data compare well to the results obtained by
the EAS-Top experiment \cite{eastopspec} and agree with extrapolations of
direct measurements to high energies \cite{aspenreview}.  Considering the
energy range above 10~GeV, at least a qualitative picture of the energy spectra
for individual mass groups emerges. The spectra seem to be compatible with
power laws with a cut-off at high energies.  The cut-off behavior indicated by
the measurements is reflected by theoretical considerations taking into account
the maximum energy attained during acceleration in supernova remnants
\cite{sveshnikova} or diffusive propagation of cosmic rays in the galactic
magnetic fields \cite{kalmykov}. 

\section{Conclusions and outlook}
The KASCADE experiment, measuring simultaneously the electromagnetic, muonic,
and hadronic shower components has reached the sensitivity to investigate
details in the development of showers in the atmosphere and to give hints for
the improvement of the description of hadronic interaction processes.  With the
KASCADE data hadronic interaction models could be improved within the last
years.  The all-particle energy spectrum is reasonably well known.  For the
first time, spectra for groups of elements could be derived in the \knee
region.  The observed spectra seem to exhibit cut-offs proportional to the
nuclear charge.  The \knee in the all-particle energy spectrum is caused by a
depression of the flux of light elements.  The increase of the mean logarithmic
mass derived from air shower observations seems to be compatible with
subsequential cut-offs for individual elements.

Energy spectra have been reconstructed with KASCADE data up to energies of
100~PeV. At these energies statistical errors start to dominate the overall
error. To improve this situation the experiment has been enlarged.  Covering an
area of 0.5~km$^2$, 37 detector stations, containing 10~m$^2$ of plastic
scintillators each, have been installed around the original KASCADE set-up
\cite{grande}.  Regular measurements with this new array and the original
KASCADE detectors, forming the KASCADE-Grande experiment, are performed since
summer 2003.  First analyses extend the lateral distributions of electrons and
muons up to 600~m \cite{glasstetterpune,vanburenpune}.  The objective is to
reconstruct primary energy spectra for elemental groups up to $10^{18}$~eV
\cite{haungsaspen}, covering the energy region of the second \knee, where the
galactic cosmic ray spectrum is expected to end \cite{aspenphen}.

A new technique to measure air showers is about to be established: The LOPES
experiment, consisting of 30 dipole antennas, operated at the site of the
KASCADE experiment, has registered first air showers, observing radio emission
in the frequency range from 40 to 80~MHz \cite{radionature}.  Analyses show a
correlation between the strength of the radio signal and the angle between the
shower axis and the direction of the geomagnetic field, indicating that the
emission process is most likely geosynchrotron emission. An increase of the
detected radio signal as function of the shower energy has been observed.  The
new technique is expected to contribute to the exploration of air showers at
energies exceeding $10^{17}$~eV through the measurement of the longitudinal
shower development.


  \vspace*{1mm}
  {\sl
  KASCADE is supported by the Ministry for Research and Education of Germany,
  the Polish State Committee for Scientific Research (KBN grant for 2004-06) and
  the Romanian National Academy for Science, Research and Technology.}


\end{document}